\documentclass[showpacs,pra,showkeys,amsmath,notitlepage]{revtex4-1}
\usepackage{color}
\definecolor{brown}{rgb}{0.8,0.6,0.3}
\definecolor{dgreen}{rgb}{0.2,0.4,0.3}

\usepackage[pdfauthor={Richard J. Mathar},pdfkeywords={GTO, ERI, 4-center},pdftoolbar=false,colorlinks=true]{hyperref}
\begin{document}
\author{Richard J. Mathar}
\homepage{http://www.mpia.de/~mathar}
\affiliation{Max-Planck Institute of Astronomy, K\"onigstuhl 17, 69117 Heidelberg, Germany}

\date{\today}
\title{Four-center Integral of a Dipolar Two-electron Potential Between $s$-type GTO's}

\begin{abstract}
We reduce two-electron 4-center products of Cartesian Gaussian Type Orbitals
with Boys' contraction to 2-center products of the form $\psi_\alpha(\mathbf{r}_i-\mathbf{A})$
$\psi_\beta(\mathbf{r}_j-\mathbf{B})$,
and compute the 6-dimensional integral over $d^3r_i d^3r_j$ over these
with the effective potential
$V_{ij} = (\mathbf{r}_i-\mathbf{r}_j)\cdot \mathbf{r}_j/|\mathbf{r}_i-\mathbf{r}_j|^3$
in terms of Shavitt's confluent hypergeometric functions.
\end{abstract}
\maketitle

\section{Format of the Integral}
In relativistic quantum chemistry, the effective electron-electron interaction
contains so-called gauge correction terms \cite{SchreckenJPC101,HelgakerJCP113,EngstromCP237}
which appear in the computation  as
energy integrals of the form
\begin{equation}
J(\alpha,\mathbf{A},
\beta,\mathbf{B},
\gamma,\mathbf{C},
\delta,\mathbf{D})
=
\int d^3r_i d^3r_j
\psi_\alpha(\mathbf{r}_i-\mathbf{A})
\psi_\beta(\mathbf{r}_i-\mathbf{B})
\frac{(\mathbf{r}_i-\mathbf{r}_j)\cdot(2\mathbf{r}_i-\mathbf{r}_j)}{|\mathbf{r}_i-\mathbf{r}_j|^3}
\psi_\gamma(\mathbf{r}_j-\mathbf{C})
\psi_\delta(\mathbf{r}_j-\mathbf{D})
\end{equation}
for orbitals $\psi$ centered at places $\mathbf A$, $\mathbf B$, $\mathbf C$ and $\mathbf D$.
The Gauss Transformation Method has been shown to calculate the integral if the 
orbitals $\psi$ are expanded in a basis of Gaussian Type Orbitals (GTO's) \cite{ShiozakiJCP138};
this manuscript basically demonstrates how dealing with the quadratic forms
in the exponentials directly also manages to reduce them to the omnipresent Confluent Hypergeometric
Functions of the electron repulsion integrals (ERI's).

\section{Reduction to 2-Center Integrals}
By the usual treatment of GTO's \cite{Boys} we contract the Gaussians related to electron $i$
and electron $j$ by defining intermediate centers
$\mathbf P$ and $\mathbf Q$:
\begin{eqnarray}
\mathbf{P}&\equiv& \frac{\alpha\mathbf{A}+\beta\mathbf{B}}{\alpha+\beta},
\label{eq.Pdef}
\\
\mathbf{Q}&\equiv& \frac{\gamma\mathbf{C}+\delta\mathbf{D}}{\gamma+\delta},
\label{eq.Qdef}
\end{eqnarray}
\begin{equation}
e^{-\alpha(\mathbf{r}-\mathbf{A})^2}
e^{-\beta(\mathbf{r}-\mathbf{B})^2}
=
\exp[-\frac{\alpha\beta}{\alpha+\beta} (\mathbf{A}-\mathbf{B})^2]
e^{-(\alpha+\beta)(\mathbf{r}-\mathbf{P})^2},
\end{equation}
\begin{equation}
e^{-\gamma(\mathbf{r}-\mathbf{C})^2}
e^{-\delta(\mathbf{r}-\mathbf{D})^2}
=
\exp[-\frac{\gamma\delta}{\gamma+\delta} (\mathbf{C}-\mathbf{D})^2]
e^{-(\gamma+\delta)(\mathbf{r}-\mathbf{Q})^2}.
\end{equation}
Prefactors of the form
$
(x_i-A_x)^{n_{x\alpha}}(y_i-A_y)^{n_{y\alpha}}(z_i-A_z)^{n_{z\alpha}}
(x_i-B_x)^{n_{x\beta}}(y_i-B_{y\beta})^{n_y}(z_i-B_z)^{n_{z\beta}}
$ introduced by Cartesian GTO's
of higher angular momentum quantum numbers $n_x$, $n_y$ and $n_z$
are also re-centered at $\mathbf{P}$ and $\mathbf{Q}$ by binomial expansion \cite{MurchieJCP26}.
Hermite Gaussians or Spherical Gaussians may be re-centered
by transformation to and from an intermediate Cartesian basis for the same goal \cite{MatharIJQC90}.

After that step the integrals are 2-center integrals:
\begin{equation}
\bar J
\equiv
\exp[-\frac{\alpha\beta}{\alpha+\beta} (\mathbf{A}-\mathbf{B})^2
-\frac{\gamma\delta}{\gamma+\delta} (\mathbf{C}-\mathbf{D})^2]
I(\kappa,\mathbf{P},\lambda,\mathbf{Q})
,
\label{eq.Jpref}
\end{equation}
where
\begin{equation}
I(\kappa,\mathbf{P},\lambda,\mathbf{Q})
\equiv
\int d^3r_i d^3r_j
\psi_\kappa(\mathbf{r}_i-\mathbf{P})
\frac{(\mathbf{r}_i-\mathbf{r}_j)\cdot(2\mathbf{r}_i-\mathbf{r}_j)}{|\mathbf{r}_i-\mathbf{r}_j|^3}
\psi_\lambda(\mathbf{r}_j-\mathbf{Q})
.
\label{eq.Idef}
\end{equation}

\section{Reduction to a Triple Integral} 
The substitution $\mathbf{R} = \mathbf{r}_i-\mathbf{r}_j$ in the integrand 
of the previous equation yields
\begin{equation}
I
=\int d^3R d^3r_j
\psi_\kappa(\mathbf{R}+\mathbf{r}_j-\mathbf{P})
\frac{\mathbf{R}\cdot(2\mathbf{R}+\mathbf{r}_j)}{R^3}
\psi_\lambda(\mathbf{r}_j-\mathbf{Q}).
\label{eq.I}
\end{equation}
\subsection{Isotropic Term}
The first term of $I=K+\bar I$ in the previous equation
is the well known 2-electron Coulomb repulsion \cite{PeterssonEJP31}:
\begin{multline}
K\equiv \int d^3R d^3r_j
\psi_\kappa(\mathbf{R}+\mathbf{r}_j-\mathbf{P})
\frac{\mathbf{R}\cdot2\mathbf{R}}{R^3}
\psi_\lambda(\mathbf{r}_j-\mathbf{Q})
=
2\int d^3R d^3r_j
\psi_\kappa(\mathbf{R}+\mathbf{r}_j-\mathbf{P})
\frac{1}{R}
\psi_\lambda(\mathbf{r}_j-\mathbf{Q})
\\
=2e^{-\kappa P^2-\lambda Q^2}\int dr_x dr_y dr_z dR_x dR_y dR_z
\\ \times
e^{-\kappa[R_x^2+R_y^2+R_z^2+r_x^2+r_y^2+r_z^2
+2R_xr_x
+2R_yr_y
+2R_zr_z
-2R_xP_x
-2R_yP_y
-2R_zP_z
-2r_xP_x
-2r_yP_y
-2r_zP_z]}
\\
\times \frac{1}{(R_x^2+R_y^2+R_z^2)^{1/2}}
e^{-\lambda[r_x^2+r_y^2+r_z^2
-2r_xQ_x
-2r_yQ_y
-2r_zQ_z]}
\\
=2e^{-\kappa P^2-\lambda Q^2}
\big[
\int dr_x dR_x
\int dr_y dR_y
\int dr_z dR_z
e^{-\kappa[R_x^2+r_x^2
+2R_xr_x
-2R_xP_x
-2r_xP_x]}
\\ \times
e^{-\kappa[R_y^2+r_y^2
+2R_yr_y
-2R_yP_y
-2r_yP_y]}
\\ \times
e^{-\kappa[R_z^2+r_z^2
+2R_zr_z
-2R_zP_z
-2r_zP_z]}
\\
\times \frac{1}{(R_x^2+R_y^2+R_z^2)^{1/2}}
e^{-\lambda[r_x^2
-2r_xQ_x]}
e^{-\lambda[r_y^2
-2r_yQ_y]}
e^{-\lambda[r_z^2
-2r_zQ_z]}
\big]
\\
=2e^{-\kappa P^2-\lambda Q^2}
\big[
\int dr_x dR_x
e^{-\kappa[R_x^2+r_x^2
+2R_xr_x
-2R_xP_x
-2r_xP_x]}
\frac{1}{(R_x^2+R_y^2+R_z^2)^{1/2}}
e^{-\lambda[r_x^2
-2r_xQ_x]}
\\ \times
\int dr_y dR_y
e^{-\kappa[R_y^2+r_y^2
+2R_yr_y
-2R_yP_y
-2r_yP_y]}
e^{-\lambda[r_y^2
-2r_yQ_y]}
\\ \times
\int dr_z dR_z
e^{-\kappa[R_z^2+r_z^2
+2R_zr_z
-2R_zP_z
-2r_zP_z]}
e^{-\lambda[r_z^2
-2r_zQ_z]}
\big]
.
\label{eq.K6}
\end{multline}

The integral over $r_z$ is handled as usual by completion of the quadratic form of $r_z$ in the
exponential
\begin{multline}
\int dr_z
e^{-\kappa[R_z^2+r_z^2
+2R_zr_z
-2R_zP_z
-2r_zP_z]}
e^{-\lambda[r_z^2
-2r_zQ_z]}
=
e^{-\kappa[R_z^2-2R_zP_z]}
\int dr_z
e^{-\kappa[r_z^2 +2R_zr_z -2r_zP_z]
-\lambda[r_z^2 -2r_zQ_z]}
\\
=
e^{-\kappa[R_z^2-2R_zP_z]}
\int dr_z
e^{-(\kappa+\lambda)r_z^2 -2\kappa R_zr_z +2\kappa r_zP_z+2\lambda r_zQ_z}
=
e^{-\kappa[R_z^2-2R_zP_z]}
\int dr_z
e^{-(\kappa+\lambda)r_z^2 -2(\kappa R_z -\kappa P_z-\lambda Q_z)r_z}
\\
=
e^{-\kappa[R_z^2-2R_zP_z]}
\int dr_z
e^{-(\kappa+\lambda)[r_z^2 +2\frac{\kappa R_z -\kappa P_z-\lambda Q_z}{\kappa+\lambda}r_z]}
\\
=
e^{-\kappa[R_z^2-2R_zP_z]}
\int dr_z
e^{-(\kappa+\lambda)[r_z^2 +2\frac{\kappa R_z -\kappa P_z-\lambda Q_z}{\kappa+\lambda}r_z
+(\frac{\kappa R_z -\kappa P_z-\lambda Q_z}{\kappa+\lambda})^2
-(\frac{\kappa R_z -\kappa P_z-\lambda Q_z}{\kappa+\lambda})^2
]
}
\\
=
e^{-\kappa[R_z^2-2R_zP_z]}
e^{-(\kappa+\lambda)[
-(\frac{\kappa R_z -\kappa P_z-\lambda Q_z}{\kappa+\lambda})^2
]}
\int dr_z
e^{-(\kappa+\lambda)[r_z^2 +2\frac{\kappa R_z -\kappa P_z-\lambda Q_z}{\kappa+\lambda}r_z
+(\frac{\kappa R_z -\kappa P_z-\lambda Q_z}{\kappa+\lambda})^2
]
}
\\
=
e^{-\kappa[R_z^2-2R_zP_z]}
e^{(\kappa+\lambda)
(\frac{\kappa R_z -\kappa P_z-\lambda Q_z}{\kappa+\lambda})^2
}
\int dr_z
e^{-(\kappa+\lambda)r_z^2
}
\\
=
e^{-\kappa[R_z^2-2R_zP_z]}
e^{
\frac{(\kappa R_z -\kappa P_z-\lambda Q_z)^2}{\kappa+\lambda}
}
\sqrt{\frac{\pi}{\kappa+\lambda}}.
\label{eq.rz}
\end{multline}
The same treatment integrates along the $r_x$ and the $r_y$ directions:
\begin{equation}
\int dr_x
e^{-\kappa[R_x^2+r_x^2
+2R_xr_x
-2R_xP_x
-2r_xP_x]}
e^{-\lambda[r_x^2
-2r_xQ_x]}
=
e^{-\kappa[R_x^2-2R_xP_x]}
e^{
\frac{(\kappa R_x -\kappa P_x-\lambda Q_x)^2}{\kappa+\lambda}
}
\sqrt{\frac{\pi}{\kappa+\lambda}}.
\label{eq.rx}
\end{equation}
\begin{equation}
\int dr_y
e^{-\kappa[R_y^2+r_y^2
+2R_yr_y
-2R_yP_y
-2r_yP_y]}
e^{-\lambda[r_y^2
-2r_yQ_y]}
=
e^{-\kappa[R_y^2-2R_yP_y]}
e^{
\frac{(\kappa R_y -\kappa P_y-\lambda Q_y)^2}{\kappa+\lambda}
}
\sqrt{\frac{\pi}{\kappa+\lambda}}.
\label{eq.ry}
\end{equation}
Insertion of the previous three equations into (\ref{eq.K6}) yields
\begin{multline}
K
=2e^{-\kappa P^2-\lambda Q^2}
(\frac{\pi}{\kappa+\lambda})^{3/2}
\int dR_x
dR_y
dR_z
e^{-\kappa[R_x^2-2R_xP_x]}
e^{
\frac{(\kappa R_x -\kappa P_x-\lambda Q_x)^2}{\kappa+\lambda}
}
\\
\times
e^{-\kappa[R_y^2-2R_yP_y]}
e^{
\frac{(\kappa R_y -\kappa P_y-\lambda Q_y)^2}{\kappa+\lambda}
}
\\
\times
e^{-\kappa[R_z^2-2R_zP_z]}
e^{
\frac{(\kappa R_z -\kappa P_z-\lambda Q_z)^2}{\kappa+\lambda}
}
\frac{1}{(R_x^2+R_y^2+R_z^2)^{1/2}}
.
\label{eq.K3}
\end{multline}
The principal axis transformation in the Gaussian exponentials with
the main variable $R_x$ is
\begin{equation}
e^{-\kappa[R_x^2-2R_xP_x]}
e^{
\frac{(\kappa R_x -\kappa P_x-\lambda Q_x)^2}{\kappa+\lambda}
}
=
\exp[\kappa P_x^2+\lambda Q_x^2]
\exp[
-\frac{\kappa\lambda}{\kappa+\lambda}
(R_x-P_x+Q_x)^2
]
.
\label{eq.paxR}
\end{equation}
Substitution of this form for $R_x$, $R_y$ and $R_z$ into (\ref{eq.K3}) produces
\begin{multline}
K
=2
e^{-\kappa P^2-\lambda Q^2}
(\frac{\pi}{\kappa+\lambda})^{3/2}
\int dR_x
dR_y
dR_z
\exp[\kappa P_x^2+\lambda Q_x^2]
\exp[
-\frac{\kappa\lambda}{\kappa+\lambda}
(R_x-P_x+Q_x)^2
]
\\
\times
\exp[\kappa P_y^2+\lambda Q_y^2]
\exp[
-\frac{\kappa\lambda}{\kappa+\lambda}
(R_y-P_y+Q_y)^2
]
\\
\times
\exp[\kappa P_z^2+\lambda Q_z^2]
\exp[
-\frac{\kappa\lambda}{\kappa+\lambda}
(R_z-P_z+Q_z)^2
]
\frac{1}{(R_x^2+R_y^2+R_z^2)^{1/2}}
\\
\\
=2
(\frac{\pi}{\kappa+\lambda})^{3/2}
\int dR_x
dR_y
dR_z
\exp[
-\frac{\kappa\lambda}{\kappa+\lambda}
(R_x-P_x+Q_x)^2
]
\\
\times
\exp[
-\frac{\kappa\lambda}{\kappa+\lambda}
(R_y-P_y+Q_y)^2
]
\\
\times
\exp[
-\frac{\kappa\lambda}{\kappa+\lambda}
(R_z-P_z+Q_z)^2
]
\frac{1}{(R_x^2+R_y^2+R_z^2)^{1/2}}
.
\label{eq.Kdiag}
\end{multline}
Definition of a new vector $\mathbf E$ and of a reduced scaling parameter $\epsilon$
\begin{equation}
\mathbf{E}\equiv \mathbf{P}-\mathbf{Q},\quad \epsilon\equiv \frac{\kappa\lambda}{\kappa+\lambda},
\label{eq.Edef}
\end{equation}
expresses (\ref{eq.Kdiag}) as an integral over the entire space of $d^3R$:
\begin{equation}
K
=2
(\frac{\pi}{\kappa+\lambda})^{3/2}
\int dR_x dR_y dR_z
\exp[ -\epsilon (\mathbf{R}-\mathbf{E})^2 ]
\frac{1}{(R_x^2+R_y^2+R_z^2)^{1/2}}.
\end{equation}
We rotate the coordinate system  such that the vector $\mathbf{E}$ points along the
polar coordinate and switch to a spherical coordinate system with radial coordinate $X$,
polar coordinate $\theta$ and azimuth $\phi$.
A factor $X^2 \sin\theta$ from the Jacobian in spherical coordinates is inserted,
\begin{multline}
K(\epsilon,\mathbf{E})
=2
(\frac{\pi}{\kappa+\lambda})^{3/2}
e^{-\epsilon E^2}
\int X^2\sin\theta dX d\theta d\phi
\exp[
-\epsilon
\{X^2-2EX\cos\theta\}
]
\frac{1}{X}
\\
=2
(\frac{\pi}{\kappa+\lambda})^{3/2}
e^{-\epsilon E^2}
\int X\sin\theta dX d\theta d\phi
\exp[
-\epsilon
\{X^2-2EX\cos\theta\}
]
\\
=2
(\frac{\pi}{\kappa+\lambda})^{3/2}
2\pi e^{-\epsilon E^2}
\int X\sin\theta dX d\theta
\exp[
-\epsilon
\{X^2-2EX\cos\theta\}
]
\\
=2
(\frac{\pi}{\kappa+\lambda})^{3/2}
2\pi e^{-\epsilon E^2}
\int X dX
\int_{-1}^1 dz
\exp[
-\epsilon
\{X^2-2EXz\}
]
\\
=2
(\frac{\pi}{\kappa+\lambda})^{3/2}
2\pi e^{-\epsilon E^2}
\int_0^\infty X dX e^{-\epsilon X^2}
\int_{-1}^1 dz
\exp[
2\epsilon EXz
]
\\
=2
(\frac{\pi}{\kappa+\lambda})^{3/2}
2\pi e^{-\epsilon E^2}
\int _0^\infty X dX e^{-\epsilon X^2}
\frac{1}{2\epsilon EX}
(
e^{2\epsilon EX}-e^{-2\epsilon EX}
)
\\
=2
(\frac{\pi}{\kappa+\lambda})^{3/2}
\frac{\pi}{\epsilon E} e^{-\epsilon E^2}
\int _0^\infty dX e^{-\epsilon X^2}
(
e^{2\epsilon EX}-e^{-2\epsilon EX}
)
\\
=2
(\frac{\pi}{\kappa+\lambda})^{3/2}
\frac{\pi}{2\epsilon^2 E^2} e^{-\epsilon E^2}
\int _0^\infty dt
e^{-t^2/(4\epsilon E^2)}
(
e^t-e^{-t}
)
=2
(\frac{\pi}{\kappa+\lambda})^{3/2}
\frac{2\pi}{\epsilon} F_0(\epsilon E^2).
\label{eq.K}
\end{multline}
The function $F_0$ is made more explicit in Appendix \ref{sec.G}.
\subsection{Dipolar Term}

The
entire focus of this manuscript is on the second term of $I$ in (\ref{eq.I}),
\begin{equation}
\bar I(\kappa,\mathbf{P},\lambda,\mathbf{Q})\equiv \int d^3R d^3r_j
\psi_\kappa(\mathbf{R}+\mathbf{r}_j-\mathbf{P})
\frac{\mathbf{R}\cdot\mathbf{r}_j}{R^3}
\psi_\lambda(\mathbf{r}_j-\mathbf{Q}).
\end{equation}

For $s$-type orbitals
along the Cartesian coordinates $\mathbf{r}_j=(r_{x},r_{y},r_{z})$
the dot product $\mathbf{R}\cdot \mathbf{r}_j$
is expanded which decomposes $\bar I$
into a sum of three contributions:
\begin{multline}
\bar I(\kappa,\mathbf{P},\lambda,\mathbf{Q})
=e^{-\kappa P^2-\lambda Q^2}\int dr_x dr_y dr_z dR_x dR_y dR_z
\\ \times
e^{-\kappa[R_x^2+R_y^2+R_z^2+r_x^2+r_y^2+r_z^2
+2R_xr_x
+2R_yr_y
+2R_zr_z
-2R_xP_x
-2R_yP_y
-2R_zP_z
-2r_xP_x
-2r_yP_y
-2r_zP_z]}
\\
\times \frac{R_xr_{x}+R_yr_{y}+R_zr_{z}}{(R_x^2+R_y^2+R_z^2)^{3/2}}
e^{-\lambda[r_x^2+r_y^2+r_z^2
-2r_xQ_x
-2r_yQ_y
-2r_zQ_z]}
\\
=e^{-\kappa P^2-\lambda Q^2}
\bigg[
\int dr_x dr_y dr_z dR_x dR_y dR_z
e^{-\kappa[R_x^2+R_y^2+R_z^2+r_x^2+r_y^2+r_z^2
+2R_xr_x
+2R_yr_y
+2R_zr_z
-2R_xP_x
-2R_yP_y
-2R_zP_z
-2r_xP_x
-2r_yP_y
-2r_zP_z]}
\\
\times \frac{R_xr_{x}}{(R_x^2+R_y^2+R_z^2)^{3/2}}
e^{-\lambda[r_x^2+r_y^2+r_z^2
-2r_xQ_x
-2r_yQ_y
-2r_zQ_z]}
\\
+
\int dr_x dr_y dr_z dR_x dR_y dR_z
e^{-\kappa[R_x^2+R_y^2+R_z^2+r_x^2+r_y^2+r_z^2
+2R_xr_x
+2R_yr_y
+2R_zr_z
-2R_xP_x
-2R_yP_y
-2R_zP_z
-2r_xP_x
-2r_yP_y
-2r_zP_z]}
\\
\times \frac{R_yr_{y}}{(R_x^2+R_y^2+R_z^2)^{3/2}}
e^{-\lambda[r_x^2+r_y^2+r_z^2
-2r_xQ_x
-2r_yQ_y
-2r_zQ_z]}
\\
+
\int dr_x dr_y dr_z dR_x dR_y dR_z
e^{-\kappa[R_x^2+R_y^2+R_z^2+r_x^2+r_y^2+r_z^2
+2R_xr_x
+2R_yr_y
+2R_zr_z
-2R_xP_x
-2R_yP_y
-2R_zP_z
-2r_xP_x
-2r_yP_y
-2r_zP_z]}
\\
\times \frac{R_zr_{z}}{(R_x^2+R_y^2+R_z^2)^{3/2}}
e^{-\lambda[r_x^2+r_y^2+r_z^2
-2r_xQ_x
-2r_yQ_y
-2r_zQ_z]}
\bigg]
\\
=e^{-\kappa P^2-\lambda Q^2}
\bigg[
\int dr_x dR_x
\int dr_y dR_y
\int dr_z dR_z
e^{-\kappa[R_x^2+r_x^2
+2R_xr_x
-2R_xP_x
-2r_xP_x]}
\\ \times
e^{-\kappa[R_y^2+r_y^2
+2R_yr_y
-2R_yP_y
-2r_yP_y]}
\\ \times
e^{-\kappa[R_z^2+r_z^2
+2R_zr_z
-2R_zP_z
-2r_zP_z]}
\\
\times \frac{R_xr_{x}}{(R_x^2+R_y^2+R_z^2)^{3/2}}
e^{-\lambda[r_x^2
-2r_xQ_x]}
e^{-\lambda[r_y^2
-2r_yQ_y]}
e^{-\lambda[r_z^2
-2r_zQ_z]}
\\
+(x\to y)+(x\to z)
\bigg]
\\
=e^{-\kappa P^2-\lambda Q^2}
\bigg[
\int dr_x dR_x
e^{-\kappa[R_x^2+r_x^2
+2R_xr_x
-2R_xP_x
-2r_xP_x]}
\frac{R_xr_{x}}{(R_x^2+R_y^2+R_z^2)^{3/2}}
e^{-\lambda[r_x^2
-2r_xQ_x]}
\\ \times
\int dr_y dR_y
e^{-\kappa[R_y^2+r_y^2
+2R_yr_y
-2R_yP_y
-2r_yP_y]}
e^{-\lambda[r_y^2
-2r_yQ_y]}
\\ \times
\int dr_z dR_z
e^{-\kappa[R_z^2+r_z^2
+2R_zr_z
-2R_zP_z
-2r_zP_z]}
e^{-\lambda[r_z^2
-2r_zQ_z]}
\\
+(x\to y)+(x\to z)
\bigg]
.
\label{eq.cart}
\end{multline}

The integrals over $r_z$ and $r_y$ are taken
from (\ref{eq.rz}) and (\ref{eq.ry}).
An additional factor $r_x$ intrudes the integrand
along the $r_x$-direction in (\ref{eq.cart}):
\begin{multline}
\int dr_x
r_x
e^{-\kappa[R_x^2+r_x^2
+2R_xr_x
-2R_xP_x
-2r_xP_x]}
e^{-\lambda[r_x^2
-2r_xQ_x]}
\\
=
e^{-\kappa[R_x^2-2R_xP_x]}
e^{-(\kappa+\lambda)[
-(\frac{\kappa R_x -\kappa P_x-\lambda Q_x}{\kappa+\lambda})^2
]}
\int dr_x
r_x
e^{-(\kappa+\lambda)[r_x^2 +2\frac{\kappa R_x -\kappa P_x-\lambda Q_x}{\kappa+\lambda}r_x
+(\frac{\kappa R_x -\kappa P_x-\lambda Q_x}{\kappa+\lambda})^2
]
}
\\
=
e^{-\kappa[R_x^2-2R_xP_x]}
e^{(\kappa+\lambda)[
(\frac{\kappa R_x -\kappa P_x-\lambda Q_x}{\kappa+\lambda})^2
]}
\int dr_x
r_x
e^{-(\kappa+\lambda)[r_x +\frac{\kappa R_x -\kappa P_x-\lambda Q_x}{\kappa+\lambda}]^2
}
\\
=
e^{-\kappa[R_x^2-2R_xP_x]}
e^{(\kappa+\lambda)[
(\frac{\kappa R_x -\kappa P_x-\lambda Q_x}{\kappa+\lambda})^2
]}
\int dt
(t
-\frac{\kappa R_x -\kappa P_x-\lambda Q_x}{\kappa+\lambda}
)
e^{-(\kappa+\lambda)t^2
}
\\
=
-\frac{\kappa R_x -\kappa P_x-\lambda Q_x}{\kappa+\lambda}
e^{-\kappa[R_x^2-2R_xP_x]}
e^{
\frac{(\kappa R_x -\kappa P_x-\lambda Q_x)^2}{\kappa+\lambda}
}
\int dt
e^{-(\kappa+\lambda)t^2
}
\\
=
-\frac{\kappa R_x -\kappa P_x-\lambda Q_x}{\kappa+\lambda}
e^{-\kappa[R_x^2-2R_xP_x]}
e^{
\frac{(\kappa R_x -\kappa P_x-\lambda Q_x)^2}{\kappa+\lambda}
]}
\sqrt{\frac{\pi}{\kappa+\lambda}}.
\end{multline}

Insertion of this equation, of (\ref{eq.ry}) and of (\ref{eq.rz})
into (\ref{eq.cart}) has reduced the 6-fold to a 3-fold integral:
\begin{multline}
\bar I
=e^{-\kappa P^2-\lambda Q^2}
\bigg[
-(\frac{\pi}{\kappa+\lambda})^{3/2}
\int dR_x
dR_y
dR_z
\frac{\kappa R_x -\kappa P_x-\lambda Q_x}{\kappa+\lambda}
e^{-\kappa[R_x^2-2R_xP_x]}
e^{
\frac{(\kappa R_x -\kappa P_x-\lambda Q_x)^2}{\kappa+\lambda}
}
\\
\times
e^{-\kappa[R_y^2-2R_yP_y]}
e^{
\frac{(\kappa R_y -\kappa P_y-\lambda Q_y)^2}{\kappa+\lambda}
}
\\
\times
e^{-\kappa[R_z^2-2R_zP_z]}
e^{
\frac{(\kappa R_z -\kappa P_z-\lambda Q_z)^2}{\kappa+\lambda}
}
\frac{R_x}{(R_x^2+R_y^2+R_z^2)^{3/2}}
\\
+(x\to y)+(x\to z)
\bigg].
\label{eq.3}
\end{multline}

\section{Reduction of the 1-particle potential}
\subsection{Quadratic Form in the Exponential}
Substitution of the form (\ref{eq.paxR}) for $R_x$, $R_y$ and $R_z$ into (\ref{eq.3}) produces
\begin{multline}
\bar I
=
e^{-\kappa P^2-\lambda Q^2}
\bigg[
-(\frac{\pi}{\kappa+\lambda})^{3/2}
\int dR_x
dR_y
dR_z
\frac{\kappa R_x -\kappa P_x-\lambda Q_x}{\kappa+\lambda}
\exp[\kappa P_x^2+\lambda Q_x^2]
\exp[
-\frac{\kappa\lambda}{\kappa+\lambda}
(R_x-P_x+Q_x)^2
]
\\
\times
\exp[\kappa P_y^2+\lambda Q_y^2]
\exp[
-\frac{\kappa\lambda}{\kappa+\lambda}
(R_y-P_y+Q_y)^2
]
\\
\times
\exp[\kappa P_z^2+\lambda Q_z^2]
\exp[
-\frac{\kappa\lambda}{\kappa+\lambda}
(R_z-P_z+Q_z)^2
]
\frac{R_x}{(R_x^2+R_y^2+R_z^2)^{3/2}}
\\
+(R_x\to R_y)+(R_x\to R_z)
\bigg]
\\
=
-(\frac{\pi}{\kappa+\lambda})^{3/2}
\int dR_x
dR_y
dR_z
\frac{\kappa R_x -\kappa P_x-\lambda Q_x}{\kappa+\lambda}
\exp[
-\frac{\kappa\lambda}{\kappa+\lambda}
(R_x-P_x+Q_x)^2
]
\\
\times
\exp[
-\frac{\kappa\lambda}{\kappa+\lambda}
(R_y-P_y+Q_y)^2
]
\\
\times
\exp[
-\frac{\kappa\lambda}{\kappa+\lambda}
(R_z-P_z+Q_z)^2
]
\frac{R_x}{(R_x^2+R_y^2+R_z^2)^{3/2}}
\\
+(R_x\to R_y)
+(R_x\to R_z)
.
\end{multline}
Along with (\ref{eq.Edef}),
\begin{multline}
\bar I
=
-(\frac{\pi}{\kappa+\lambda})^{3/2}
\int dR_x dR_y dR_z
\frac{\kappa R_x -\kappa P_x-\lambda Q_x}{\kappa+\lambda}
\exp[ -\frac{\kappa\lambda}{\kappa+\lambda} (\mathbf{R}-\mathbf{E})^2 ]
\frac{R_x}{(R_x^2+R_y^2+R_z^2)^{3/2}}
\\
+(R_x\to R_y)
+(R_x\to R_z)
\\
=
-(\frac{\pi}{\kappa+\lambda})^{3/2}
\int dR_x dR_y dR_z
\frac{\kappa \mathbf{R}\cdot \mathbf{R} -(\kappa \mathbf{P}+\lambda \mathbf{Q})\cdot \mathbf{R}}{\kappa+\lambda}
\exp[ -\epsilon (\mathbf{R}-\mathbf{E})^2 ]
\frac{1}{(R_x^2+R_y^2+R_z^2)^{3/2}}
.
\end{multline}

The exponent in this integrand
involves the cosine of the angle
between the vectors $\mathbf{R}$ and $\mathbf{E}$,
\begin{equation}
\exp[ -\epsilon (\mathbf{R}-\mathbf{E})^2 ]
=
\exp[
-\epsilon
\{R^2+E^2-2ER(\sin\theta\sin\theta_E\cos(\phi-\phi_E)+\cos\theta\cos\theta_E)\}
]
\end{equation}
where polar and azimuthal angles are defined
in the usual manner:
\begin{eqnarray}
\mathbf{R}=R(\sin\theta\cos\phi, \sin\theta\sin\phi,\cos\theta);
\\
\mathbf{E}=E(\sin\theta_E\cos\phi_E, \sin\theta_E\sin\phi_E,\cos\theta_E).
\label{eq.Epolar}
\end{eqnarray}

As the only noticeable idea in this calculation, the inverse coordinate transformation
(\ref{eq.Om}) rotates the $R$-coordinate system such that the polar axis of the new coordinate system
points towards $\mathbf E$, so the cosine in the dot product
$\mathbf{R}\cdot \mathbf{E}$ is just the cosine of the polar coordinate of $\mathbf X$ in the new coordinate system
observed in (\ref{eq.omega}):
\begin{equation}
\mathbf{R}=\mathbf{\Omega}^{-1}\mathbf{X};
\end{equation}
\begin{equation}
\mathbf{R}\cdot \mathbf{E}=\mathbf{\Omega}^{-1}\mathbf{X}\cdot \mathbf{E} = \mathbf{X}\cdot \mathbf{\Omega}\mathbf{E}.
\end{equation}
\begin{equation}
\mathbf{\Omega}^{-1}=
\left(
\begin{array}{rrr}
(1-\cos \theta_E)\sin^2\phi_E+\cos\theta_E & -(1-\cos\theta_E)\sin\phi_E\cos\phi_E & \sin\theta_E\cos\phi_E\\
-(1-\cos\theta_E)\cos\phi_E\sin\phi_E & (1-\cos \theta_E)\cos^2\phi_E+\cos\theta_E & \sin\theta_E\sin\phi_E\\
-\sin\theta_E\cos\phi_E & -\sin\theta_E\sin\phi_E& \cos\theta_E \\
\end{array}
\right).
\end{equation}
The rotation preserves lengths,
$R=|\mathbf{R}|
=X=|\mathbf{X}|
$.
\begin{equation}
\bar I
=
-(\frac{\pi}{\kappa+\lambda})^{3/2}
\int dX_x dX_y dX_z
\frac{\kappa \mathbf{X}\cdot \mathbf{X} -(\kappa \mathbf{P}+\lambda \mathbf{Q})\cdot \mathbf{\Omega}^{-1}\mathbf{X}}{\kappa+\lambda}
\exp[ -\epsilon (\mathbf{\Omega}^{-1}\mathbf{X}-\mathbf{E})^2 ]
\frac{1}{(X_x^2+X_y^2+X_z^2)^{3/2}}
\end{equation}
\begin{equation}
=
-(\frac{\pi}{\kappa+\lambda})^{3/2}
\int dX_x dX_y dX_z
(\frac{\kappa}{\kappa+\lambda} X^2-\mathbf{E}'\cdot \mathbf{\Omega}^{-1}\mathbf{X})
\exp[ -\epsilon (\mathbf{\Omega}^{-1}\mathbf{X}-\mathbf{E})^2 ]
\frac{1}{(X_x^2+X_y^2+X_z^2)^{3/2}}
\end{equation}
\begin{equation}
=
-(\frac{\pi}{\kappa+\lambda})^{3/2}
[\frac{\kappa}{\kappa+\lambda}\bar I_1
-\bar I_2
],
\label{eq.Isplit}
\end{equation}
where we have defined  the vector $\mathbf E'$ via
\begin{equation}
\mathbf{E}'\equiv \frac{\kappa\mathbf{P}+\lambda\mathbf{Q}}{\kappa+\lambda}.
\label{eq.Eprdef}
\end{equation}

\subsection{Isotropic Part}
The term $\bar I_1$ in (\ref{eq.Isplit}) involves a factor $X^2 \sin\theta$ from the Jacobian in spherical coordinates,
a factor $\mathbf{X}\cdot \mathbf{X}=X^2$ from the dot product, and the dipolar $X^3$ in the denominator:
\begin{multline}
\bar I_{1}(\epsilon,\mathbf{E})
=
\int dX_x dX_y dX_z
X^2
\exp[ -\epsilon (\mathbf{X}-\mathbf{E})^2 ]
\frac{1}{(X_x^2+X_y^2+X_z^2)^{3/2}}
\\
=
e^{-\epsilon E^2}
\int X^2\sin\theta dX d\theta d\phi
X^2
\exp[
-\epsilon
\{X^2-2EX\cos\theta\}
]
\frac{1}{(X_x^2+X_y^2+X_z^2)^{3/2}}
\\
=
e^{-\epsilon E^2}
\int X\sin\theta dX d\theta d\phi
\exp[
-\epsilon
\{X^2-2EX\cos\theta\}
]
\\
=
2\pi e^{-\epsilon E^2}
\int X\sin\theta dX d\theta
\exp[
-\epsilon
\{X^2-2EX\cos\theta\}
]
\\
=
2\pi e^{-\epsilon E^2}
\int X dX
\int_{-1}^1 dz
\exp[
-\epsilon
\{X^2-2EXz\}
]
\\
=
2\pi e^{-\epsilon E^2}
\int_0^\infty X dX e^{-\epsilon X^2}
\int_{-1}^1 dz
\exp[
2\epsilon EXz
]
\\
=
2\pi e^{-\epsilon E^2}
\int _0^\infty X dX e^{-\epsilon X^2}
\frac{1}{2\epsilon EX}
(
e^{2\epsilon EX}-e^{-2\epsilon EX}
)
\\
=
\frac{\pi}{\epsilon E} e^{-\epsilon E^2}
\int _0^\infty dX e^{-\epsilon X^2}
(
e^{2\epsilon EX}-e^{-2\epsilon EX}
)
\\
=
\frac{\pi}{2\epsilon^2 E^2} e^{-\epsilon E^2}
\int _0^\infty dt
e^{-t^2/(4\epsilon E^2)}
(
e^t-e^{-t}
)
=
\frac{2\pi}{\epsilon} F_0(\epsilon E^2).
\label{eq.barI1}
\end{multline}
The function $F_0$ is made more explicit in Appendix \ref{sec.G}.
The gradient with respect to $\mathbf E$ is an application of (\ref{eq.Fdiff}):
\begin{equation}
\nabla_{\mathbf E} \bar I_1
= -4\pi F_1(\epsilon E^2) \mathbf{E}.
\end{equation}
This indicates that working out the integrals for 4-center orbitals
of Cartesian Gaussians
beyond the $(0,0,0)$-triple of ``orbital'' quantum numbers are tractable
through
repeated differentiation with respect to the locations of the four centers \cite{HuzinagaPTPS40,FlockeJCC29}.

\subsection{Dipolar Part}
In the other integral of (\ref{eq.Isplit}),
$$
\bar I_2 = \int d^3X \mathbf{E}' \cdot \mathbf{\Omega}^{-1}\mathbf{X}
\exp[-\epsilon(\mathbf{\Omega}^{-1}\mathbf{X}-\mathbf{E})^2]
\frac{1}{X^3}
,
$$
we compute three components defined by moving the $\mathbf{\Omega}$
operator to the
vector $\mathbf{E}'$:
\begin{equation}
\mathbf{E}'
\cdot
\mathbf{\Omega}^{-1}\mathbf{X}
=
\mathbf{\Omega}
\mathbf{E}'
\cdot
\mathbf{X}
=
H_xX_x+H_yX_y+H_zX_z
\end{equation}
where we have defined the vector $\mathbf H\equiv
\mathbf{\Omega}
\mathbf{E}'
$.
\begin{equation}
\bar I_2 = H_x\bar I_{2x}+H_y\bar I_{2y}+H_z\bar I_{2z}.
\label{eq.I2xyz}
\end{equation}
Its $z$-component is obtained with (\ref{eq.Om}):
\begin{equation}
H_z = \frac{1}{E} \mathbf{E}\cdot \mathbf{E'}.
\label{eq.Hz}
\end{equation}

The integrals $\bar I_{2x}$ and $\bar I_{2y}$ vanish while integrating over the azimuth $\phi$:
\begin{multline}
\bar I_{2x}
=
\int dX_x dX_y dX_z
X_x
\exp[ -\frac{\kappa\lambda}{\kappa+\lambda} (\mathbf{\Omega}^{-1}\mathbf{X}-\mathbf{E})^2 ]
\frac{1}{(X_x^2+X_y^2+X_z^2)^{3/2}}
\\
=
e^{-\epsilon E^2}
\int dX \sin \theta d\theta d\phi
\cos\phi\sin\theta
\exp[
-\epsilon
\{X^2-2EX\cos\theta\}
]
=0
.
\end{multline}
\begin{multline}
\bar I_{2y}
=
\int dX_x dX_y dX_z
X_y
\exp[ -\frac{\kappa\lambda}{\kappa+\lambda} (\mathbf{\Omega}^{-1}\mathbf{X}-\mathbf{E})^2 ]
\frac{1}{(X_x^2+X_y^2+X_z^2)^{3/2}}
\\
=
e^{-\epsilon E^2}
\int dX \sin \theta d\theta d\phi
\sin\phi\sin\theta
\exp[
-\epsilon
\{X^2-2EX\cos\theta\}
]
=0
.
\end{multline}
So the only finite contribution to (\ref{eq.I2xyz})
is from the component coupled to $H_z$:
\begin{multline}
\bar I_{2z}(\epsilon,\mathbf E)
=
\int dX_x dX_y dX_z
X_z
\exp[ -\frac{\kappa\lambda}{\kappa+\lambda} (\mathbf{\Omega}^{-1}\mathbf{X}-\mathbf{E})^2 ]
\frac{1}{(X_x^2+X_y^2+X_z^2)^{3/2}}
\\
=
e^{-\epsilon E^2}
\int dX \sin \theta d\theta d\phi
\cos\theta
\exp[
-\epsilon
\{X^2-2EX\cos\theta\}
]
\\
=
2\pi e^{-\epsilon E^2}
\int dX \sin \theta d\theta
\cos\theta
\exp[
-\epsilon
\{X^2-2EX\cos\theta\}
]
\\
=
2\pi e^{-\epsilon E^2}
\int_0^\infty dX \int_{-1}^1 dt
t
\exp[
-\epsilon
\{X^2-2EXt\}
]
\\
=
2\pi e^{-\epsilon E^2}
\int_0^\infty dX
e^{-\epsilon X^2}
\int_{-1}^1 dt
t
\exp[ 2\epsilon EXt ]
\\
=
2\pi e^{-\epsilon E^2}
\int_0^\infty dX
e^{-\epsilon X^2}
\frac{1}{(2\epsilon EX)^2}
[
e^{2\epsilon EX}(2\epsilon EX-1)
+
e^{-2\epsilon EX}(2\epsilon EX+1)
]
\\
=
2\pi e^{-\epsilon E^2}
\frac{1}{2\epsilon E}
\int_0^\infty dt
e^{-t^2/(4\epsilon E^2)}
\frac{1}{t^2}
[
e^t(t-1)
+
e^{-t}(t+1)
]
=4\pi E F_1(\epsilon E^2).
\label{eq.Fdef}
\end{multline}

The auxiliary special function $F_1$ is computed via the error function in Appendix \ref{sec.F}.
The gradient with respect to $\mathbf E$ is an application of (\ref{eq.Fdiff})
and of the product rule of differentiation:
\begin{equation}
\nabla_{\mathbf E} \bar I_{2z}
=
\frac{4\pi}{E} [F_1(\epsilon E^2)
-
2\epsilon E^2 F_2(\epsilon E^2)
]\mathbf{E}
.
\end{equation}

\section{Summary}
In numerical practise the steps of obtaining $J$ are:
\begin{enumerate}
\item Define the intermediate centers $\mathbf{P}$ and $\mathbf{Q}$ with their effective
scaling factors $\kappa+\beta$ and $\lambda+\delta$ via (\ref{eq.Pdef}) and (\ref{eq.Qdef});
\item
Calculate the exponential pre-factor in (\ref{eq.Jpref});
\item
Implement Shavitt's functions $F_0$ and $F_1$ for positive real-valued arguments;
\item
Calculate the contribution $K$ with (\ref{eq.K});
\item
Calculate the contribution $\bar I$ from (\ref{eq.Isplit}):
\begin{enumerate}
\item Calculate the two vectors $\mathbf{E}$, $\mathbf{E}'$ and parameter $\epsilon$ in (\ref{eq.Edef}) and (\ref{eq.Eprdef});
\item Calculate $\bar I_2=H_z\bar I_{2z}$ as the product of (\ref{eq.Hz}) and (\ref{eq.Fdef}).
\item Calculate $\bar I_1$ in (\ref{eq.barI1}).
\end{enumerate}
\item
Calculate (\ref{eq.Idef})
\begin{equation}
I=K+\bar I
=
\frac{2\pi}{\epsilon}(\frac{\pi}{\kappa+\lambda})^{3/2}
\left[
\frac{\kappa+2\lambda}{\kappa+\lambda}
F_0(\epsilon E^2)
+2\epsilon \mathbf{E}\cdot\mathbf{E}' F_1(\epsilon E^2)
\right]
.
\end{equation}
\end{enumerate}

\appendix
\section{Coordinate rotation}

The orthogonal unimodular $3\times 3$ matrix which rotates points
by an angle $\theta_E$
around the right-handed axis with Cartesian coordinates
$(\omega_1,\omega_2,\omega_3)$, normalized to unit length $\omega_1^2+\omega_2^2+\omega_3^2=1$,
is \cite{BalakrishnanR4}\cite[(2.21)]{Morawiec}
\begin{equation}
\mathbf{\Omega}=
\left(
\begin{array}{rrr}
(1-\cos \theta_E)\omega_1^2+\cos\theta_E & (1-\cos\theta_E)\omega_1\omega_2-\sin\theta_E\omega_3 & (1-\cos\theta_E)\omega_1\omega_3+\sin\theta_E\omega_2\\
(1-\cos\theta_E)\omega_1\omega_2+\sin\theta_E\omega_3 & (1-\cos \theta_E)\omega_2^2+\cos\theta_E & (1-\cos\theta_E)\omega_2\omega_3-\sin\theta_E\omega_1\\
(1-\cos\theta_E)\omega_1\omega_3-\sin\theta_E\omega_2 & (1-\cos\theta_E)\omega_2\omega_3+\sin\theta_E\omega_1 & (1-\cos \theta_E)\omega_3^2+\cos\theta_E \\
\end{array}
\right).
\label{eq.Omega}
\end{equation}
We wish to find the axis that rotates the Cartesian vector (\ref{eq.Epolar})
to the image $E(0,0,1)$, such that
\begin{equation}
\mathbf{\Omega}
\cdot \left(\begin{array}{c}
\cos\phi_E \sin\theta_E\\
\sin\phi_E \sin\theta_E\\
\cos\theta_E\\
\end{array}\right)=\left(\begin{array}{c}0\\0\\1\end{array}\right).
\label{eq.omega}
\end{equation}
The rotation axis
is
the cross product between the point in space and its image:
\begin{equation}
\left(\begin{array}{c}
\cos\phi_E \sin\theta_E\\
\sin\phi_E \sin\theta_E\\
\cos\theta_E\\
\end{array}\right)\times
\left(\begin{array}{c}0\\0\\1\end{array}\right)
=
\left(\begin{array}{c}
\sin\phi_E\sin\theta_E\\
-\cos\phi_E\sin\theta_E\\
0
\end{array}\right)
.
\end{equation}
Normalized to unit length it constructs the axis vector $\mathbf\omega$ with Cartesian components
\begin{equation}
\left(\begin{array}{c}
\omega_1\\
\omega_2\\
\omega_3
\end{array}\right)
=
\left(\begin{array}{c}
\sin\phi_E\\
-\cos\phi_E\\
0
\end{array}\right).
\end{equation}
Insertion of these three components into (\ref{eq.Omega}) yields the rotation matrix
applicable to (\ref{eq.omega}):
\begin{equation}
\mathbf{\Omega}
=
\left(
\begin{array}{rrr}
(1-\cos \theta_E)\sin^2\phi_E+\cos\theta_E & -(1-\cos\theta_E)\sin\phi_E\cos\phi_E & -\sin\theta_E\cos\phi_E\\
-(1-\cos\theta_E)\cos\phi_E\sin\phi_E & (1-\cos \theta_E)\cos^2\phi_E+\cos\theta_E & -\sin\theta_E\sin\phi_E\\
\sin\theta_E\cos\phi_E & \sin\theta_E\sin\phi_E& \cos\theta_E \\
\end{array}
\right).
\label{eq.Om}
\end{equation}
The inverse rotation is represented by the inverse matrix
(which equals the transpose matrix) and established through
the substitution $\theta_E\to -\theta_E$.

\section{Auxiliary Integral $F_0$}\label{sec.G}
The radial integral (\ref{eq.K}) resp.\ (\ref{eq.barI1}) is solved by Taylor Expansion of the $\sinh t$, followed by the substitution
$t^2=s$ and integration over $s$ with
\cite[3.351.3]{GR}
\begin{equation}
\int_0^\infty e^{-s/k}s^n ds = n!k^{n+1}.
\label{eq.intexp}
\end{equation}
\begin{multline}
F_0(k)\equiv
\frac{1}{4k}e^{-k}\int_0^\infty dt e^{-t^2/(4k)}(e^t-e^{-t})
=
\frac{1}{4k}e^{-k}\int_0^\infty dt e^{-t^2/(4k)}
2\sum_{l=1,3,5,\ldots} \frac{t^l}{l!}
\\
=
\frac{1}{4k}e^{-k}\int_0^\infty ds e^{-s/(4k)}\sum_{l\ge 0}\frac{s^l}{(2l+1)!}
\\
=
\frac{1}{4k}e^{-k}\sum_{l\ge 0}\frac{l! }{(2l+1)!}(4k)^{l+1}
\\
=
e^{-k}
\sum_{l\ge 0}
\frac{[\Gamma(l+1)]^2}{\Gamma(2l+2)}
\frac{(4k)^l}{l!}
.
\label{eq.F0lser}
\end{multline}
Application of the duplication formula for the $\Gamma$-function \cite[6.1.18]{AS}
and rewriting the $\Gamma$-functions as Pochhammer symbols converts the series
to a Confluent Hypergeometric Function \cite{RoyAMM94,SlaterHyp}:
\begin{equation}
F_0(k)
=
e^{-k}
{}_1F_1(1;3/2;k)
=
{}_1F_1(1/2;3/2;-k)
=
\frac{1}{2}k^{-1/2}\gamma(1/2,k)
=
\frac{1}{2}k^{-1/2}\sqrt{\pi}\mathrm{erf}(\surd k).
\label{eq.Gerf}
\end{equation}
For small arguments the Taylor expansion is \cite[7.1.5]{AS}
\begin{equation}
F_0(k) \stackrel{k\to 0}{\longrightarrow}
1-\frac{1}{3}k+\frac{1}{10}k^2-\frac{1}{42}k^3+\frac{1}{108}k^4+\cdots
.
\end{equation}

\section{Auxiliary Integral $F_1$}\label{sec.F}

The auxiliary function introduced in (\ref{eq.Fdef}) for real-valued argument
$k\ge 0$ turns out to be closely related to
the error function \cite{GautschiCiteseer}.
Very similar to the calculation in Appendix \ref{sec.G}, the exponentials
in the integral that depend linearly on $t$ are expanded in Taylor series
\cite[1.212]{GR},
summation and integration are interchanged,
and integration via (\ref{eq.intexp}) yields
a Confluent Hypergeometric Series:
\begin{multline}
F_1(k)
=
\frac{1}{4k}e^{-k}\int_0^\infty dt 
e^{ -t^2/(4k)}
\frac{1}{t^2}
[e^{-t}(1+t)+e^t(t-1)]
\\
=
\frac{1}{4k}e^{-k}\int_0^\infty dt 
e^{ -t^2/(4k)}
2\sum_{l\ge 0}
t^{2l+1}
\frac{2l+2}{(2l+3)!}
\\
=
\frac{1}{4k}e^{-k}\int_0^\infty ds 
e^{ -s/(4k)}
\sum_{l\ge 0}
s^l
\frac{2l+2}{(2l+3)!}
\\
=
\frac{1}{4k}e^{-k}
\sum_{l\ge 0}
(4k)^{l+1}
\frac{(2l+2)l!}{(2l+3)!}
\\
=
2e^{-k}
\sum_{l\ge 0}
\frac{(l+1)!l!}{(2l+3)!}
\frac{(4k)^l}{l!}
=
\frac{1}{3}e^{-k}
{}_1F_1(1;5/2;k).
\label{eq.F1lser}
\end{multline}
Kummer's transformation \cite[9.212]{GR} and a succession of well-known formulas
for the Incomplete Gamma-function \cite[13.1.27,13.6.10,6.5.22]{AS}
rephrase $F_1$ in terms of the error function:
\begin{equation}
F_1(k)=
\frac{1}{3}
{}_1F_1(3/2;5/2;-k)
=
\frac{1}{2}k^{-3/2}\left[\frac{\sqrt{\pi}}{2}\mathrm{erf}(\sqrt{k})-\sqrt{k}e^{-k}\right].
\label{eq.Ferf}
\end{equation}
For small arguments \cite[7.1.5]{AS}
\begin{equation}
F_1(k) \stackrel{k\to 0}{\longrightarrow}
\frac{1}{3}-\frac{1}{5}k+\frac{1}{14}k^2-\frac{1}{54}k^3+\frac{1}{264}k^4+\cdots
.
\end{equation}

\section{Shavitt's $F$-integral}
$F_0$ and $F_1$ are special cases of Shavitt's $F_\nu$-functions
\cite{ShavittJCP43,TaketaJPSP21,NiukkanenIJQC18,RysJCC4,Dupuis2001,SagarIJQC42,MatharNumAlg36}
\begin{equation}
F_\nu(t)\equiv \int_0^1 u^{2\nu}e^{-tu^2}du
=\frac{1}{2\nu+1}{}_1F_1(\nu+\frac12;\nu+\frac32;-t)
=\frac{1}{2\nu+1}e^{-t}{}_1F_1(1;\nu+\frac32;t).
\end{equation}
Its first derivative is
\begin{equation}
\frac{d}{dt}F_\nu(t) = - F_{\nu+1}(t).
\label{eq.Fdiff}
\end{equation}
The recurrence of the Confluent Hypergeometric Function \cite[13.4.7]{AS}
\begin{equation}
b(1-b+z){}_1F_1(a;b;z)+
b(b-1){}_1F_1(a-1;b-1;z)-
az{}_1F_1(a+1;b+1;z)=0
\end{equation}
establishes through insertion of $a=\nu+3/2$, $b=\nu+5/2$ the equivalent
\begin{equation}
z F_{\nu+2}(z)
-(z+\nu+3/2) F_{\nu+1}(z)
+(\nu+1/2) F_\nu(z)
=0
.
\end{equation}
The Laplace transform is
\begin{equation}
\hat F_\nu(s)\equiv \int_0^\infty e^{-st} F_\nu(t)dt
=
\int_0^1 \frac{1}{s+u^2} du u^{2\nu},
\end{equation}
with recurrence
\begin{equation}
\hat F_{\nu+1}(s)
=
\frac{1}{2\nu+1}-s\hat F_{\nu}(s),
\end{equation}
starting at
\begin{equation}
\hat F_0(s)
= \int_0^1\frac{1}{s+u^2}du
= \frac{1}{\surd s} \arctan \frac{1}{\surd s}.
\end{equation}
The only singularity of $\hat F_{\nu}(s)$ is at $s=0$.

By performing the analysis of (\ref{eq.F0lser}) or (\ref{eq.F1lser}) backwards we find for general integer $\nu$
\begin{multline}
F_\nu(k) =
\frac{1}{2\nu+1}e^{-k}\sum_{l\ge 0}
\frac{(1)_l k^l}{(\nu+3/2)_l l!}
\\
=
\frac{e^{-k}}{4k}\frac{2}{2\nu+1}
\int_0^{\infty} dt e^{-t^2/(4k)} t
{}_0F_1(-;\nu+3/2;t^2/4)
\\
=
\frac{e^{-k}}{4k} (2\nu-1)!! 2\sum_{l\ge 0}
\int_0^{\infty} dt e^{-t^2/(4k)}\frac{1}{t^{2\nu}} t^{2\nu+2l+1}
\frac{(2l+2)(2l+4)\cdots (2l+2\nu)}{(2\nu+1+2l)!}
\\
=
(2\nu-1)!! \frac{e^{-k}}{4k} 
\int_0^{\infty} dt e^{-t^2/(4k)}\frac{1}{t^{2\nu}}
\left[e^{-t}(\beta_{\nu,0}+\beta_{\nu,1}t+\beta_{\nu,2}t^2+\beta_{\nu,3}t^3+\cdots)-(t\to -t)\right]
\label{eq.eli}
.
\end{multline}
\begin{table}
\caption{Coefficients $\beta_{\nu,l}$ in (\ref{eq.eli}) in row $\nu$ and column $l$.
$\beta_{\nu,l}=0$ if $l>\nu$, above the diagonal.
}
\begin{ruledtabular}
\begin{tabular}{r|rrrrrrrrrr}
 & 0 & 1 & 2 & 3 & 4 &5 &6 &7\\
\hline
0 & -1 & \\
1 & 1 & 1 & \\
2 & -3 & -3 & -1 & \\
3 & 15 & 15 & 6 & 1 & \\
4 & -105 & -105 & -45 & -10 & -1 & \\
5 & 945 & 945 & 420 & 105 & 15 & 1 & \\
6 & -10395 & -10395 & -4725 & -1260 & -210 & -21 & -1 & \\
7 & 135135 & 135135 & 62370 & 17325 & 3150 & 378 & 28 & 1 & \\
\end{tabular}
\end{ruledtabular}
\label{tab.alp}
\end{table}
The absolute values of the matrix elements $\beta_{\nu,l}$ are Sequence A001497
in the Online Encyclopedia
of Integer Sequences \cite{EIS},
illustrated in Table \ref{tab.alp}.
The row $\nu=0$ in the table represents (\ref{eq.F0lser}), the row $\nu=1$ represents (\ref{eq.F1lser}).
The closed form
\begin{equation}
\beta_{\nu,l}
=\left\{
\begin{array}{ll}
(-1)^{\nu+1} 2^{l-\nu}\frac{\nu!}{l!}\binom{2\nu-l}{\nu}, & 0\le l \le \nu; \\
0, & \mathrm{else}.
\end{array}
\right.
\end{equation}
is readily available \cite{EIS}.
The matrix inverse of $\beta$ is also a lower triangular array with elements
essentially obtained by transposition of $\beta$ itself:
\begin{equation}
(\beta^{-1})_{\nu,l} =\left\{
\begin{array}{ll}
(-1)^{\nu +1}|\beta_{1+l,1-\nu+2l}| ,&  \lfloor \nu/2\rfloor \le l \le \nu; \\
0, & \mathrm{else}.
\end{array}
\right.
\end{equation}

\bibliography{all}

\end{document}